\documentclass[aps,prl,twocolumn,superscriptaddress,notitlepage,nofootinbib,showkeys,showpacs]{revtex4-1}
\pdfoutput=1
\usepackage{graphicx}
\usepackage[dvipsnames]{xcolor}
\usepackage{amsmath}
\usepackage {cancel}
\allowdisplaybreaks
\usepackage{amssymb}

\usepackage{slashed}
\usepackage[utf8]{inputenc}
\usepackage[colorlinks=true,linkcolor=blue,bookmarksopen,bookmarksnumbered]{hyperref}
\usepackage{color}
\usepackage[normalem]{ulem}
\usepackage{soul}
\usepackage{units}
\usepackage{rotating}
\usepackage{hhline,multirow,tabularx}
\usepackage{bm}
\usepackage{hyperref}
\usepackage[export]{adjustbox}
\usepackage{mdframed}
\usepackage{comment}
\usepackage{natbib}
\usepackage{etoolbox}

\def\bea#1\eea{\begin{align}#1\end{align}}

\newcommand{\bef}{\begin{figure}[htb]\centering}
\newcommand{\eef}{\end{figure}}

\newcommand{\nn}{\nonumber}

\newcommand{\shao}[1]{\marginpar{\footnotesize\textbf{SHAO}}}

\def\<{\langle}
\def\>{\rangle}

\def\cos{\hbox{cos}}
\def\sin{\hbox{sin}}

\makeatletter
\patchcmd\frontmatter@PACS@format{\addvspace{11\p@}}{}{}{}
\pretocmd\frontmatter@keys@format{\addvspace{11\p@}}{}{}
\patchcmd{\titleblock@produce}
  {\@pacs@produce\@pacs\@keywords@produce\@keywords}
  {\@keywords@produce\@keywords\@pacs@produce\@pacs}
  {}{}
\makeatother

\usepackage{lipsum}
\begin{document}

\title{Spin asymmetry and dipole moments in $\tau$-pair production with ultraperipheral heavy ion collisions}

\author{Ding Yu Shao}
\affiliation{Department of Physics, Center for Field Theory and Particle Physics, Fudan University, Shanghai, 200433, China}
\affiliation{Key Laboratory of
Nuclear Physics and Ion-beam Application (MOE), Fudan University, Shanghai, 200433, China}
\affiliation{Shanghai Research Center for Theoretical Nuclear Physics, National Natural Science Foundation of China (NSFC) and Fudan University, Shanghai 200438, China}

\author{Bin Yan}
\email{yanbin@ihep.ac.cn (corresponding author)}
\affiliation{Institute of High Energy Physics, Chinese Academy of Sciences, Beijing 100049, China}

\author{Shu-Run Yuan}
\affiliation{School of Physics, Peking University, Beijing 100871, China}

\author{Cheng Zhang}
\affiliation{Department of Physics, Center for Field Theory and Particle Physics, Fudan University, Shanghai, 200433, China}


\begin{abstract}
The anomalous magnetic (MDM) and electric (EDM) dipole moments of the $\tau$ lepton serve as crucial indicators of new physics beyond the Standard Model. Leveraging azimuthal angular asymmetry as a novel tool in ultraperipheral collisions (UPCs), we attain unparalleled precision in the study of these key properties. Driven by the highly linear polarization of coherent photons, this method uniquely enables both the MDM and EDM to contribute to the $\cos2\phi$ angular distribution in similar magnitudes. Importantly, our approach substantially narrows the parameter space, excluding more than half of it compared to expected UPC-based measurements reliant solely on the total cross-section. This method not only provides improved constraints but also minimizes the need for additional theoretical assumptions, and offers a novel avenue to probe the EDM effects.
\end{abstract}



\maketitle

\section{ Introduction}

The precise measurements of the anomalous magnetic (MDM) and electric (EDM) dipole moments of leptons are crucial for providing stringent tests for the Standard Model (SM) and offering potential insights into new physics (NP) phenomena~\cite{Czarnecki:2001pv,Giudice:2012ms,Kurz:2014wya,Kurz:2015bia,Kurz:2016bau,Liu:2018xkx,Liu:2020qgx,Aebischer:2021uvt,Li:2021koa,Cirigliano:2021peb,Workman:2022ynf,Cao:2022htd,Wen:2023xxc,Cao:2023juc}. In particular, the EDMs provide an additional source of {\it CP} violation beyond the CKM mechanism, which could potentially contribute to an explanation for the universe's matter-antimatter asymmetry problem~\cite{Sakharov:1967dj}. The sensitivity of the MDM and EDM to NP effects is expected to be proportional to the masses of the fermions involved~\cite{Giudice:2012ms}, as a result, the $\tau$ lepton, with its relatively large mass, is anticipated to be the most sensitive lepton for detecting NP effects through MDM and EDM measurements. Furthermore, the recent exciting news from the Fermilab muon MDM measurement revealed a significant deviation between experimental data and SM predictions at a significance level of $5\sigma$~\cite{Muong-2:2023cdq}, which also highlights the importance and urgency of probing the MDM and EDM of the $\tau$ lepton, as it may provide crucial insights into the underlying physics responsible for this deviation.

However, a significant challenge arises due to the short lifetime of the $\tau$ lepton, which prevents the feasibility of directly  measuring its MDM and EDM using the techniques similar to those used for the electron and muon. Instead, the information about the $\tau$ lepton's MDM and EDM can only be obtained from the measurements of $\tau$ production and decays at colliders~\cite{delAguila:1991rm,DELPHI:2003nah,Bernabeu:2007rr,Atag:2010ja,Billur:2013rva,Eidelman:2016aih,Chen:2018cxt,Fu:2019utm,Beresford:2019gww,Dyndal:2020yen,Verducci:2023cgx}. Notably, the most precise measurements of MDM come from the DELPHI Collaboration, which used $\gamma\gamma\to\tau^+\tau^-$ production at the Large Electron Positron (LEP) collider, {\it i.e.} $a_\tau=-0.018\pm 0.017$~\cite{DELPHI:2003nah}, while the most stringent limits on the EDM are obtained from the measurement of {\it CP}-odd observables in the $e^+e^-\to \tau^+\tau^-$ process by the Belle Collaboration~\cite{Belle:2002nla}, {\it i.e.} ${\rm Re}(d_\tau)=(1.5\pm 1.7)\times 10^{-17} {\rm e\cdot cm}$ and ${\rm Im}(d_\tau)=(-0.83\pm 0.86)\times 10^{-17} {\rm e\cdot cm}$, where $a_\tau$ and $d_\tau$ respectively represent the MDM and EDM of the $\tau$ lepton and are described by the following effective $\tau^+\tau^-\gamma$ vertex~\footnote{The anapole moment of tau lepton cannot contribute to lepton pair production in UPCs, as the photons involved are quasi-real.},
\begin{align}\label{eq:formfact}
\Gamma_{\rm eff.}^\mu(q^2)= -i e \left [i F_2(q^2) +F_3(q^2) \gamma^5 \right]  \frac{\sigma^{\mu \nu}q_\nu}{2 m_\tau}.
\end{align}
Here the $q$ represents the momentum of the off-shell photon.
In the limit of $q^2\to 0$, the functions $F_2(0)=a_\tau$ and $F_3(0)=2m_\tau d_\tau/e$, where $m_\tau$ is the mass of the $\tau$ lepton. It is important to note that the leading contribution to $a_\tau$ in SM is the one-loop Schwinger term $\alpha_e/2\pi\simeq 0.00116$~\cite{Schwinger:1948iu}, which is still one order of magnitude smaller than the expected sensitivity in current experimental measurements. Additionally, it also shows that the $d_\tau\sim\mathcal{O}(10^{-37})~{\rm e\cdot cm}$ in SM, which is far from the experimental sensitivity~\cite{Yamaguchi:2020eub,Yamaguchi:2020dsy}. As a result, it would be reasonable to neglect their contribution in SM and consider $a_\tau$ and $d_\tau$ as a contribution from NP in this study.

\begin{figure}\centering
    \includegraphics[scale=0.4]{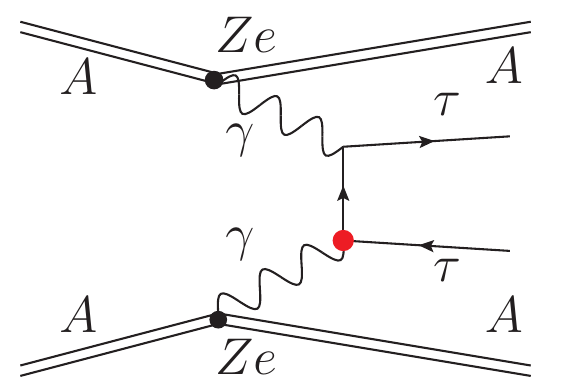}
    \caption{Pair production of $\tau$ leptons from ultraperipheral heavy ion collisions. The red dot denotes the effective couplings from MDM and EDM.}
    \label{fig:Feyndiag}
\end{figure}

In addition to lepton collider measurements, ultraperipheral heavy ion collisions (UPCs) provide an alternative avenue for probing the $a_\tau$ and $d_\tau$, as well as other NP effects because the electromagnetic interactions will be dominant for the di-lepton production with a large impact parameter, defined as $b>2R_A$ where $R_A$ is the ion radius, and the production cross section can be much enhanced by the photon flux with a factor of $Z^4$, where $Z$ is the atomic number~\cite{delAguila:1991rm,Knapen:2016moh,Beresford:2019gww,Dyndal:2020yen,Xu:2022qme,Verducci:2023cgx} ($Z=82$ for lead). Recently, the observation of $\tau$-pair production in UPCs has been reported by both the ATLAS and CMS collaborations~\cite{ATLAS:2022ryk,CMS:2022arf}. These measurements have placed strong constraints on the $a_\tau$ and $d_\tau$ for a single parameter analysis, with an accuracy that is competitive with constraints obtained from lepton colliders~\cite{DELPHI:2003nah}.

However, to date, all the knowledge regarding the $a_\tau$ and $d_\tau$ relies heavily  on the theoretical assumptions made in the analysis, such as the frequent exclusion of other possible NP effects, which could potentially affect the accuracy and reliability of the extracted values from the data.
To address this issue, in this Letter we propose to utilize the spin asymmetry of $\tau$-pair production in UPCs to probe the $a_\tau$ and $d_\tau$ simultaneously. The representative Feynman diagram at the leading order is shown in Fig.~\ref{fig:Feyndiag}, and the diagram illustrates that the $a_\tau$ and $d_\tau$ would be the leading NP effects impacting $\tau$-pair production in UPCs.  Previous studies have demonstrated that coherent photons in UPCs exhibit highly linear polarization, with their polarization vectors aligned with the direction of their transverse momenta, which are typically around $k_\perp\sim 30~{\rm MeV}$ (corresponding to the inverse of the nuclear radius)~\cite{Li:2019sin,Li:2019yzy,Ma:2023dac}. Recently, this property of the coherent photon has been confirmed by the STAR Collaboration at  Relativistic Heavy Ion Collider (RHIC) by observing a significant $\cos 4\phi$ azimuthal asymmetry in di-electron production~\cite{STAR:2019wlg} with results consistent with theoretical predictions~\cite{Li:2019yzy,Li:2019sin}, where $\phi$ is the azimuthal angle between total and relative momenta of the di-electron. This breakthrough presents a new opportunity to utilize spin asymmetry in UPCs to explore NP effects.

In this Letter, we will demonstrate that both the $a_\tau$ and $d_\tau$ uniquely contribute to the $\cos2\phi$ angular modulation with similar magnitudes. Utilizing both the azimuthal asymmetry and cross section measurements in $\tau$-pair production in UPCs, we can simultaneously constrain both $a_\tau$ and $d_\tau$ with comparable or higher accuracy to other methods in the literature. Importantly, our approach requires fewer theoretical assumptions. Moreover, by utilizing spin asymmetry, we significantly constrain the parameter space for $a_\tau$ and $d_\tau$, excluding over half of it compared to using cross-section measurements alone.

\section{Theoretical formalism}

The $\tau$-pair production in UPCs occurs via the photon-photon fusion process, {\it i.e.,} $\gamma(x_1 P+ k_{1\perp})+\gamma(x_2\bar{P}+ k_{2\perp}) \rightarrow \tau^{+}\left(p_{1}\right)+\tau^{-}\left(p_{2}\right)$.
Previous studies have shown that the dominant contribution to the azimuthal asymmetry due to the coherent photons comes from the low total transverse momentum of the lepton pair ($\bm q_{\perp} \equiv \bm p_{1 \perp}+\bm p_{2 \perp}$)~\cite{Li:2019sin,Li:2019yzy,STAR:2019wlg,Xiao:2020ddm,Klein:2020jom,Zhao:2022dac,Brandenburg:2022tna,Zha:2018tlq,Wang:2021kxm,Wang:2022gkd}. Therefore, our analysis in this work will focus on the kinematic region where the $\tau$ leptons are produced nearly back-to-back, satisfying the condition  $q_\perp = |\bm q_\perp|\ll p_{1\perp},p_{2\perp}$. It is imperative to recognize that when incorporating impact parameter dependence in our cross-section analysis, we are obliged to extend beyond traditional Transverse Momentum Dependent (TMD) factorization. This extension is necessitated by differing transverse momenta in the incoming photons between the amplitude and its conjugate.~\cite{Vidovic:1992ik,Hencken:1994my,Krauss:1997vr}.

Following the formalism developed in Refs.~\cite{Vidovic:1992ik,Hencken:1994my,Krauss:1997vr}, the joint impact parameter $b_\perp$ and $q_\perp$ dependent cross section from the SM and dipole interactions can be cast into the form~\footnote{Any additional $\cos(n\phi)$ angular modulations from final state QED radiations \cite{Li:2019yzy,Hatta:2021jcd,Shao:2022stc,Shao:2023zge} have been consciously disregarded due to its impact for our results is negligible.},
\begin{align}\label{eq:born}
& \frac{\mathrm{d} \sigma}{\mathrm{d}^{2} \bm q_{\perp} \mathrm{d}^{2} \bm P_{\perp} \mathrm{d} y_{1} \mathrm{d} y_{2} \mathrm{d}^{2} \bm b_{\perp}} \!=\! \frac{\alpha_e^2}{2M^4\pi^2} \!\left[A_0+B_0^{(1)}F_2+B_0^{(2)} F_2^2 \right.\nn\\
 &\left.+C_0^{(2)} F_3^2+\left(A_2+B_2^{(2)}F_2^2+C_2^{(2)}F_3^2\right) \cos 2 \phi+A_4\cos 4\phi \right],
\end{align}
where $\bm P_{\perp} \equiv (\bm p_{1 \perp}-\bm p_{2 \perp})/2$, $y_{1,2}$ are rapidities of the leptons, $M$ is the invariant mass of the lepton pair, $\alpha_e$ is the fine-structure constant, and $\phi$ is the angle between transverse momentum $\bm q_{\perp}$ and $\bm P_{\perp}$. In this equation, the terms $A_0,B_0^{(1),(2)},C_0^{(2)}$ and $A_{2,4},B_2^{(2)},C_2^{(2)}$ encapsulate the unpolarized differential cross section and the azmimuthal asymmetry arising from linearly polarized coherent photons, respectively. The coefficients $A_{0,2,4}$ correspond to the contributions from the SM, while the other terms arise from the contributions of $a_\tau$ and $d_\tau$. Owing to the orthogonality of the polarization vectors of photons with different helicity states, both the $a_\tau$ and $d_\tau$ can only generate the angular modulations as shown in Eq.~\eqref{eq:born}. In the limit of $m_\tau\ll P_\perp$, the coefficients in Eq.~\eqref{eq:born} can be written as,
\begin{align}
A_{0}\! & = \frac{M^2-2P_\perp^2}{P_{\perp}^{2} }  \!{\cal \int}[{\mathrm{d}\cal K}_\perp]\cos(\phi_{k_{1\perp}}\!\!-\phi_{\bar k_{1\perp}}\!\!+\phi_{k_{2\perp}}\!\!-\phi_{\bar k_{2\perp}}), \label{eq:A0QED}\\
A_2\! & = \frac{8m_\tau^{2}}{P_{\perp}^{2}}{\cal \int}[{d\cal K}_\perp] \cos\left(\phi_{k_{1\perp}}-\phi_{k_{2\perp}}\right)\notag \\
&\hspace{2.3cm} \times\cos(\phi_{\bar k_{1\perp}}+\phi_{\bar k_{2\perp}}-2\phi_{q_\perp}),\label{eq:A2QED}\\
A_4\! &= -2{\cal \int}[{d\cal K}_\perp] \cos(\phi_{k_{1\perp}}\!\!+\phi_{\bar k_{1\perp}}\!\!+\phi_{k_{2\perp}}\!\!+\phi_{\bar k_{2\perp}}-4\phi_{q_\perp}),
\label{eq:A4QED}\\
B_{0}^{(1)}\! & = \frac{4 M^2}{P_{\perp}^{2} }  \!{\cal \int}[{\mathrm{d}\cal K}_\perp]\sin(\phi_{k_{1\perp}}\!\!-\phi_{\bar k_{2\perp}})\sin(\phi_{\bar k_{1\perp}}\!\!-\phi_{k_{2\perp}}), \label{eq:B0QED}\\
B_{0}^{(2)}\! & = C_0^{(2)} = \frac{2 M^2}{m_{\tau}^{2}}  \!{\cal \int}[{\mathrm{d}\cal K}_\perp]\cos(\phi_{k_{1\perp}}\!\!-\phi_{\bar k_{1\perp}}) \notag \\
&\hspace{2.3cm} \times \cos(\phi_{k_{2\perp}}\!\!-\phi_{\bar k_{2\perp}}), \label{eq:B0'QED}\\
B_{2}^{(2)}\! & = C_2^{(2)} = -\frac{2 M^2}{m_{\tau}^{2}}  \!{\cal \int}[{\mathrm{d}\cal K}_\perp]
\cos(\phi_{k_{1\perp}}-\phi_{k_{2\perp}})\notag \\
&\hspace{2.3cm} \times \cos(\phi_{\bar k_{1\perp}}+\phi_{\bar k_{2\perp}}-2\phi_{q_\perp}). \label{eq:B2QED}
\end{align}
The lepton mass correction and the QED resummation effects for the SM have been discussed in Refs.~\cite{Hatta:2021jcd,Shao:2022stc,Shao:2023zge}.
We should note that the equality $B_0^{(2)}=C_0^{(2)}$ is only correct when we ignore the correction of $m_\tau$. The full $m_\tau$ dependence will be included in the following numerical analysis. The shorthanded notation for the integration is defined as follows,
\begin{align}
{\cal \int}&[\mathrm{d}{\cal K}_\perp] \equiv \int \mathrm{d}^{2}\bm k_{1\perp}\mathrm{d}^{2}\bm k_{2\perp}\mathrm{d}^{2}\bar {\bm k}_{1\perp}\mathrm{d}^{2}\bar {\bm k}_{2\perp}e^{i(\bm k_{1\perp}-\bar {\bm k}_{1\perp})\cdot \bm b_{\perp}} \notag \\
&\times \delta^{(2)}(\bm k_{1\perp}+\bm k_{2\perp}-\bm q_{\perp}) \delta^{(2)}(\bar {\bm k}_{1\perp}+\bar {\bm k}_{2\perp}-{\bm q}_{\perp}) \notag \\
&\times \mathcal{F}(x_1,{\bm k}_{1\perp}^{2})\mathcal{F}(x_2, {\bm k}_{2\perp}^{2})\mathcal{F}(x_1, \bar {\bm k}_{1\perp}^{2})\mathcal{F}(x_2, \bar {\bm k}_{2\perp}^{2}),
\end{align}
where the transverse momenta of the initial photons in the amplitude and its conjugate amplitude are denoted by $\bm k_{1,2\perp}$ and $\bar {\bm k}_{1,2\perp}$, respectively. The corresponding azimuthal angles of the initial photons are denoted as $\phi_{k_{1,2\perp}}$ and $\phi_{\bar k_{1,2\perp}}$. $x_1$ and $x_2$ represent the longitudinal momentum fractions of the incoming photons,
\begin{align}\label{eq:energy_frac}
x_{1,2} &= \frac{\sqrt{\bm P_{\perp}^2+m_\tau^2}}{\sqrt{S}}\left(e^{\pm y_{1}}+e^{\pm y_{2}}\right),
\end{align}
with $S$ is the center of mass energy. Using the equivalent photon approximation, a well-established method for computing UPC observable, the photon’s probability amplitude function ${\cal F}(x,{\bm k}_\perp^2)$ is defined as~\cite{Bertulani:1987tz,Vidovic:1992ik},
\begin{eqnarray}\label{eq:gamma_amp}
    {\cal F}(x,{\bm k}_\perp^2)=\frac{Z \sqrt{\alpha_e}}{\pi} |{\bm k}_\perp|
   \frac{F({\bm k}_\perp^2+x^2M_p^2)}{({\bm k}_\perp^2+x^2M_p^2)}.
\end{eqnarray}
Here $M_p$ is the proton mass. The nuclear charge density distribution in momentum space is taken from the STARlight generator~\cite{Klein:2016yzr},
\begin{eqnarray}
F(\kappa^2)=\frac{3[\sin(\kappa R_{A})-\kappa R_{A}\cos( R_{A})]}{( \kappa R_{A})^{3}(a^{2}\kappa^{2}+1)},
\label{eq:NCDD}
\end{eqnarray}
where $a=0.7$ fm and $R_{A}=1.1A^{1/3}$ fm with $A=208$ for the isotope used at the LHC. Such a parametrization is very close to the Fourier transform of the Woods-Saxon distribution numerically~\cite{Li:2019sin,Li:2019yzy}.

As shown in $A_2$ from Eq.~\eqref{eq:A2QED}, the leading order QED effect for the $\cos2\phi$ modulation is suppressed by the lepton mass, rendering the $\cos2\phi$ asymmetry in electron-positron pair production negligible. However, this effect becomes significant in muon and tauon production, as confirmed by the STAR collaboration~\cite{Zhou:2022gbh}. Notably, a distinct $\cos2\phi$ modulation arising from the MDM and EDM is observed in the coefficients $B_2^{(2)}$ and $C_2^{(2)}$ from Eq.~\eqref{eq:B2QED}. This modulation is primarily determined by the kinematic factors associated with the chirality flip of the dipole interactions,  {\it i.e.} the coefficients $B_2^{(2)}=C_2^{(2)}\propto M^2/m_\tau^2$ as shown in Eq.~\eqref{eq:B2QED}. Consequently, a hard kinematic phase space region can enhance the contribution of MDM and EDM to the $\cos2\phi$ modulation.

\section{Phenomenology}

We now present the projected azimuthal distribution of $\tau$-pair production from the MDM and EDM in ultra-peripheral Pb+Pb collisions at the LHC with a center-of-mass energy \(\sqrt{S} = 5.02\) TeV. To ensure a valid theoretical prediction, we implement a transverse impact parameter constraint with \( b_\perp > 2R_{\text{WS}}\), where \(R_{\text{WS}}\) represents the Wood-Saxon radius of the Pb nucleus, taken to be $6.68$ fm.

To extract the coefficient of the $\cos2\phi$ modulation, which is sensitive to the $a_\tau$ and $d_\tau$, we can define the azimuthal asymmetry of $\tau$ pair production based on the $\cos 2\phi$ distribution as follows,
\begin{eqnarray}
    A_{c2\phi}=\frac{\sigma(\cos 2\phi>0)-\sigma(\cos 2\phi<0)}{\sigma(\cos 2\phi>0)+\sigma(\cos 2\phi<0)}\,,
    \label{eq:asy}
\end{eqnarray}
where $\sigma(\cos 2\phi>0)$ and $\sigma(\cos 2\phi<0)$ represent the integrated cross section with $\cos 2\phi>0$ and $\cos 2\phi<0$, respectively. It is evident that this construction ensures that the large coefficient $A_4$ in Eq.~\eqref{eq:born} from the SM does not contribute to the asymmetry, and the contribution from the $A_2$ will be suppressed by $m_\tau$ due to the ratio $A_2/B_2^{(2)}\propto 4m_\tau^4/(P_\perp^2M^2)$.

\begin{figure}[t]\centering
    \includegraphics[scale=0.5]{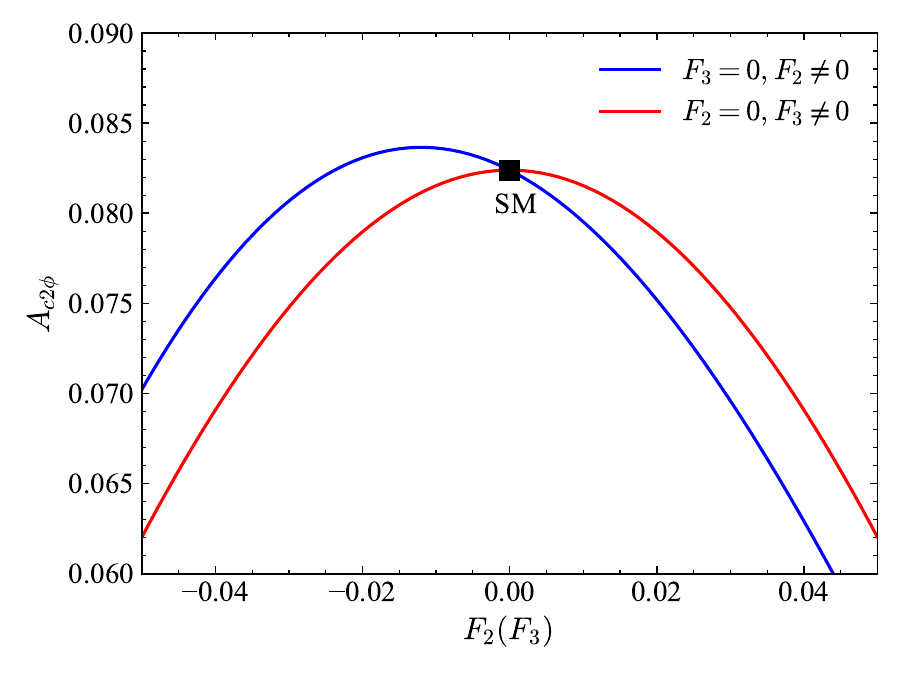}
    \caption{Theoretical predictions for the azimuthal asymmetry of $\tau^+ \tau^-$ production in ultra-peripheral Pb$+$Pb collisions with $\sqrt{S}=5.02~\text{TeV}$ and $p_T^\tau > 3~\text{GeV}$. The blue and red curves represent the prediction of $F_2=a_\tau$ and $F_3=2m_\tau d_\tau/e$, respectively.}
    \label{fig:asy}
\end{figure}

In Fig.~\ref{fig:asy}, we display the azimuthal asymmetry $A_{c2\phi}$ from the MDM alone (blue line) and EDM alone (red line) with a requirement of $p_T^\tau>3~{\rm GeV}$. It is evident that both $a_\tau$ and $d_\tau$ can significantly modify the asymmetry $A_{c2\phi}$ in UPCs. Therefore this new spin asymmetry can provide a sensitive probe for the $a_\tau$ and $d_\tau$.

However, it is important to consider the reconstruction of the $\tau$ leptons from their decay products when investigating the experimental sensitivity to $a_\tau$ and $d_\tau$ through $\tau$-pair production in UPCs. The primary decay channels of the $\tau$ lepton include leptonic decay with one charged lepton, and hadronic decay with one or three charged hadrons (pions or kaons). The experimental measurements have considered the following typical event topologies for the signals: (a) one muon and one electron; (b) one muon and one charged hadron; and (c) one muon and three charged hadrons. Using a data sample of one muon and three charged hadrons collected from 5.02 TeV Pb+Pb collisions, with an integrated luminosity of 404 ${ \mu b^{-1}}$, the CMS collaboration obtained the fiducial cross section of $\tau$-pair production $\sigma = 4.8 \pm 0.6 (\rm{stat}) \pm 0.5 (\rm{syst}) \, \mu b$~\cite{CMS:2022arf}. On the other hand, the ATLAS collaboration considered all three aforementioned signals with an integrated luminosity of 1.44 ${\rm nb}^{-1}$ and obtained a signal strength of $\mu_{\tau\tau}=1.03\pm 0.05 ({\rm stat})\pm 0.03 ({\rm syst})$~\cite{ATLAS:2022ryk}.

To ensure consistency with the fiducial phase space of the experiments, we investigate the impact of kinematic cuts on the reconstructed $\tau$ distribution from $\tau$-pair production in UPCs. We utilize gamma-UPC~\cite{Shao:2022cly} with MadGraph5~\cite{Alwall:2014hca} to generate events of $\tau$ pair in ultraperipheral collisions involving lead nuclei, and then employ PYTHIA 8~\cite{Bierlich:2022pfr}  for decay, showering and hadronization. By applying the same basic kinematic cuts used by the ATLAS and CMS collaborations, we obtain a distribution of final states similar to that observed by the ATLAS and CMS, with the exception of some discrepancies in the low $p_T$ region of muon, which may be attributed to unapplied cuts and detector effects. Subsequently, we compare the $p_T$ distribution of the $\tau$ lepton before and after implementing the cuts implemented by ATLAS and CMS, and determine the cut efficiencies as a function of $p_T^\tau$. We assume that linear photon polarization has negligible impact on the kinematic distributions of the $\tau$ decay products. As a result, we can convolute the differential cross section of $\tau$-pair production in UPCs, as given in Eq.~\eqref{eq:born}, with the cut efficiencies obtained from our simulation to mimic the fiducial phase space selection employed in the experiments.

\begin{figure}
    \includegraphics[width=0.45\textwidth]{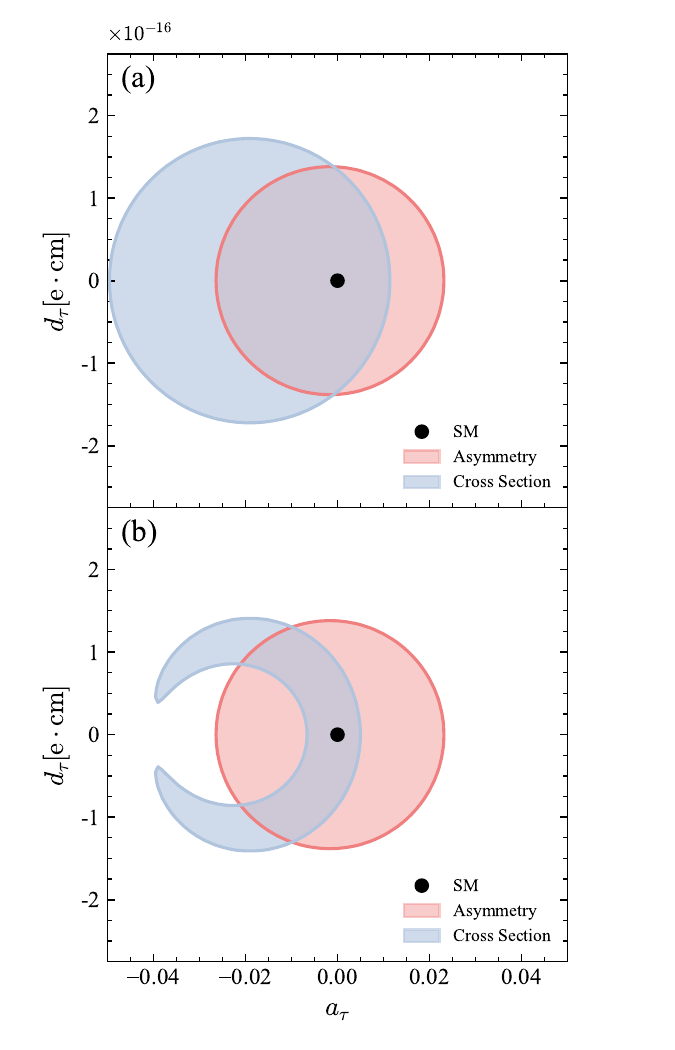}
\caption{Expected sensitivity of $a_\tau$ and $d_\tau$ at 68\% C.L. in ultraperipheral Pb+Pb collisions. The results are based on an expected integrated luminosity of $30 \, \text{nb}^{-1}$ and a center-of-mass energy $\sqrt{S} = 5.02 \, \text{TeV}$. The blue and red bands represent constraints from the cross section and azimuthal asymmetry of $\tau$ pair production, respectively. Figure (a) assumes the same systematic uncertainties for the cross section measurements as those reported by the ATLAS~\cite{ATLAS:2022ryk} and CMS~\cite{CMS:2022arf} collaborations. Figure (b) assumes a reduced systematic uncertainty of 1\%.
}
\label{fig:F23limit}
\end{figure}

Now we present the projected constraints of the azimuthal asymmetry in Eq.~\eqref{eq:asy} on the $a_\tau$ and $d_\tau$ in UPCs. The statistical uncertainty of $A_{c2\phi}$ is given by
\begin{eqnarray}
    \delta A_{c2\phi} = \sqrt{\frac{1-(A_{c2\phi})^2}{\sigma \cdot \mathcal{L}}} \simeq \frac{1}{\sqrt{\sigma \cdot \mathcal{L}}} \simeq \frac{\delta \sigma_{stat}}{\sigma},
\end{eqnarray}
where the second equality holds due to the small value of $A_{c2\phi}$ in the SM, and the third equality takes into account the fact that the statistical uncertainty of event number $N$ is $\sqrt{N}$. We assume that the systematic uncertainties cancel out in $A_{c2\phi}$ and can be ignored in this study. To estimate the sensitivity for probing the $a_\tau$ and $d_\tau$ parameters using $A_{c2\phi}$, we assume that the cut efficiencies for future Pb+Pb collision would be same as the current values of the ATLAS and CMS experiments, and the error $\delta A_{c2\phi}$ can be obtained by properly rescaling for any integrated luminosity based on the current experimental analysis. However, the systematic uncertainties in total cross section measurements usually cannot be ignored, and we consider two cases in this study. The first case assumes that the systematic uncertainties are the same as the current values for the ATLAS and CMS experiments, {\it i.e.,} 3\% for the ATLAS and 10\% for the CMS measurements. The second case assumes a systematic uncertainty of 1\% for both the ATLAS and CMS experiments. By conducting the pseudo experiments, we perform a combined $\chi^2$ analysis given by,
\begin{eqnarray}
    \chi^2 = \sum_i\left[ \frac{V^i-V_{SM}^i}{\delta V^i}\right]^2\,,
\end{eqnarray}
where $V^i$ with $i={\rm ATLAS, CMS}$ represents the total fiducial cross section or azimuthal asymmetry, and $\delta V^i$ corresponds to the total uncertainty of the $i$-th measurement. For future Pb+Pb collision with an integrated luminosity of $30~{\rm nb}^{-1}$  and a center-of-mass energy $\sqrt{S}=5.02\,\rm{TeV}$, we obtain the expected constraints on the $a_\tau$ and $d_\tau$ at 68\% confidence level (C.L.) from the total cross section (blue band) and $A_{c2\phi}$ (red band) in Fig.~\ref{fig:F23limit} with (a) and (b) correspond to the two scenarios of systematic uncertainty of the cross section mentioned earlier~\footnote{It has been shown that a template fit to the muon transverse-momentum distribution from $\tau$-lepton candidates could further improve the constraint on $a_\tau$ and $d_\tau$ because the systematic uncertainties could be well controlled by the  dimuon ($\gamma\gamma\to\mu^+\mu^-$) process~\cite{ATLAS:2022ryk}. However, the impact of this technique is beyond the scope of this work and is not considered in our analysis.}. As can be observed, incorporating the azimuthal asymmetry into the analysis can significantly reduce the parameter space of $a_\tau$ and $d_\tau$, and potentially resolve possible degeneracies that may arise when considering only the total cross section. Assuming that the systematic uncertainty can be optimized as shown in Fig.~\ref{fig:F23limit} (b), the prospective 68\% C.L. constraints for $a_{\tau}$ and $d_{\tau}$ are $-0.02 < a_{\tau} < 0.005$ and $|d_{\tau}| < 1.2 \times 10^{-16} \, {\rm e\cdot cm}$. It is important to note that the different cut efficiencies for the ATLAS and CMS measurements will cause a shift in the center of the allowed parameter space of $(a_\tau,d_\tau)$ between the ATLAS and CMS cross section measurements. As a result, when we combine the results from the ATLAS and CMS collaborations, we obtain a crescent shape in the $(a_\tau,d_\tau)$ plane, as shown by the blue band in Fig.~\ref{fig:F23limit}(b).

Before closing this section, we would like to emphasize that the cut efficiencies used in this study are not optimized based on the spin asymmetry. With the inclusion of more decay modes of $\tau$ leptons and further optimization, we expect that future experimental analyses could significantly improve the limit for $d_\tau$. We anticipate that this limit could reach the order of $\mathcal{O}(10^{-17})~{\rm e \cdot cm}$ or even smaller.

\section{Conclusions}

In this Letter, we have introduced a groundbreaking methodology that leverages the azimuthal asymmetry in UPCs for a high-precision investigation of the anomalous magnetic and electric dipole moments of the $\tau$ lepton. Our approach significantly refines the azimuthal $\cos2\phi$ modulation in photon-photon fusion processes due to the unique properties of coherent photons in UPCs. Utilizing pseudo-data based on measurements from the ATLAS and CMS collaborations, as well as current decay-product cut efficiencies, we have successfully demonstrated that our method imposes robust constraints on both the MDM and EDM of the $\tau$ lepton simultaneously. A key advantage of our approach is its minimal reliance on theoretical assumptions, significantly reducing the impact of potential new physics effects that often complicate analysis in traditional methods, thus increasing the robustness of the results. The unique and highly linear polarization of coherent photons in UPCs offers an unparalleled setting for scrutinizing {\it CP} violation mechanisms via EDM. Given these advances, we are optimistic that our approach will not only stimulate new measurements at the LHC but also inspire fresh directions in the quest for physics beyond the Standard Model.

\section{Acknowledgments}
The authors thank Qing-Hong Cao, Mei-Sen Gao and Yandong Liu for helpful discussion. Ding Yu Shao and Cheng Zhang are supported by the National Science Foundations of China under Grant No.~12275052 and No.~12147101 and the Shanghai Natural Science Foundation under Grant No.~21ZR1406100.
Bin Yan is supported by the IHEP under Contract No. E25153U1.
Shu-Run Yuan is supported in part by the National Science Foundation of China under Grants No.11725520, No.11675002 and No.12235001.

\bibliography{ref}

\onecolumngrid
\newpage
\appendix

 \section*{Supplemental material}

 In the supplemental section, we present a comprehensive derivation of the cross section dependent on both the impact parameter $b_\perp$ and transverse momentum imbalance $q_\perp$ for photon-induced tau pair production in the ultraperipheral heavy ion collisions, denoted as $A_{1}A_{2}\rightarrow A_{1}A_{2}\tau^{+}\tau^{-}$. This process is characterized by the effective vertex $\tau^+\tau^-\gamma$, which can be represented by the following equation
 \begin{align}\label{eq:eff_int}
 \Gamma^\mu(q^2)= -i e \left\{\gamma^{\mu} F_1(q^2) + \frac{\sigma^{\mu \nu}q_\nu}{2 m_\tau} \left [i F_2(q^2) +F_3(q^2) \gamma^5 \right] \right\},
 \end{align}
where $q$ is the momentum of the off-shell photon, and the form factors $F_1$, $F_2$, and $F_3$ are functions of $q^2$. In the limit as $q^2$ approaches zero, these form factors simplify to $F_1(0)=1$, $F_2(0)=a_\tau$, and $F_3(0)=2m_\tau d_\tau/e$.

We apply the equivalent photon approximation to describe the electromagnetic fields of a moving nucleus as a flux of quasi-real photons. This approach allows us to approximate the electromagnetic production cross section of $\tau^+\tau^-$ pairs in nuclear collisions by the cross section for the process $\gamma(k_{1})\gamma(k_{2})\rightarrow \tau^{+}(p_{1})\tau^{-}(p_{2})$, incorporating the equivalent photon distributions. Following the methodology established in Refs.~\cite{Vidovic:1992ik,Hencken:1994my,Krauss:1997vr}, the cross section can be expressed as
\begin{align}\label{eq:xsec-def}
\sigma  =  \int\frac{\mathrm{d}^{2}\bm b_{\perp}\mathrm{d}^{3}\bm p_{1}\mathrm{d}^{3}\bm p_{2}}{(2\pi)^{3}2E_{1}(2\pi)^{3}2E_{2}}\biggl|\int\frac{d^{4}k_{1}d^{4}k_{2}}{(2\pi)^{4}(2\pi)^{4}}(2\pi)^{4}\delta^{(4)}(k_{1}+k_{2}-p_{1}-p_{2})\mathcal{M}_{\mu\nu}(k_{1},k_{2},p_{1},p_{2})A_{1}^{\mu}(k_{1},b_{\perp})A_{2}^{\nu}(k_{2},0)\biggr|^{2},
\end{align}
where $\bm p_{1,2}$ and $E_{1,2}$ represent the three-momenta and energies of the tau leptons, respectively. The electromagnetic potentials $A_{1}^{\mu}$ and $A_{2}^{\mu}$ are given by
\begin{align}
&A_{1}^{\mu}(k_{1}, b_{\perp})=2\pi Ze\frac{F(-k_{1}^{2})}{-k_{1}^{2}}\delta(k_{1}\cdot u_{1})u_{1}^{\mu}e^{i\bm k_{1\perp}\cdot \bm b_{\perp}}, \\
&A_{2}^{\mu}(k_{2},0)=2\pi Ze\frac{F(-k_{2}^{2})}{-k_{2}^{2}}\delta(k_{2}\cdot u_{2})u_{2}^{\mu}.
\end{align}
These equations take into account the nuclear charge density distribution $F(-k^2)$, as defined in Eq.~\eqref{eq:NCDD}, and the velocities of the heavy ions, denoted as $u_{1,2}^{\mu}=\gamma_L(1,0,0,\pm v)$, where $\gamma_L$ is the Lorentz contraction factor. Furthermore, the vertex function describing the $\gamma(k_{1})\gamma(k_{2})\rightarrow \tau^{+}(p_{1})\tau^{-}(p_{2})$ process is represented by
\begin{align}
\mathcal{M}_{\mu \nu}=e^{2}\bar{u}(p_{1})\left[\Gamma_{\mu}\frac{\cancel{p}_{1}-\cancel{k}_{1}+m_\tau}{(p_{1}-k_{1})^{2}-m_\tau^{2}}\Gamma_{\nu}+\Gamma_{\nu}\frac{\cancel{p}_{1}-\cancel{k}_{2}+m_\tau}{(p_{1}-k_{2})^{2}-m_\tau^{2}}\Gamma_{\mu}\right]v(p_{2}).
\end{align}
Substituting the expressions of the electromagnetic potential $A^\mu$ into the cross section as defined in \eqref{eq:xsec-def}, we arrive at the following equation
\begin{align}
\sigma = & \frac{Z^{4}e^{4}}{4v^{2}\gamma_L^{4}}\int\frac{\mathrm{d}^{2}\boldsymbol{b}_{\perp}\mathrm{d}^{3}\boldsymbol{p}_{1}\mathrm{d}^{3}\boldsymbol{p}_{2}}{(2\pi)^{2}2E_{1}2E_{2}} \notag \\
&\times \biggl|\int\frac{\mathrm{d}^{2}\boldsymbol{k}_{1\perp}\mathrm{d}^{2}\boldsymbol{k}_{2\perp}}{(2\pi)^{2}(2\pi)^{2}}\delta^{(2)}(\boldsymbol{k}_{1\perp}+\boldsymbol{k}_{2\perp}-\boldsymbol{p}_{1\perp}-\boldsymbol{p}_{2\perp})\mathcal{M}_{\mu\nu}(k_{1},k_{2},p_{1},p_{2})e^{i\bm k_{1\perp}\cdot \bm b_{\perp}}u_{1}^{\mu}u_{2}^{\nu}\frac{F(-k_{1}^{2})}{-k_{1}^{2}}\frac{F(-k_{2}^{2})}{-k_{2}^{2}}\biggr|^{2},
\end{align}
where we have utilized the delta function $\delta(k_i \cdot u_i)$ present in the electromagnetic potential $A^\mu$ to integrate out the longitudinal and energy components of the momenta, $k_{1,2}^3$ and $k_{1,2}^0$.

As demonstrated in Ref.~\cite{Vidovic:1992ik}, by applying certain approximations that reflect the nature of the virtual photons emanating from the electromagnetic fields, we can simplify the product of the four-velocities and the matrix element. Specifically, these approximations allow us to represent $u_{1}^\mu u_{2}^{\nu} \mathcal{M}_{\mu \nu}\left(k_1, k_2,p_1,p_2\right)$ as a product involving the transverse momenta and the energies of the virtual photons, yielding
\begin{align}
u_{1}^\mu u_{2}^{\nu} \mathcal{M}_{\mu \nu}\left(k_1, k_2,p_1,p_2\right) \approx \gamma_L^2 \frac{\bm k_{1\perp}^i}{k^0_1} \frac{\bm k_{2 \perp}^j}{k_2^0} \mathcal{M}_{i j}\left(k_1,k_2, p_1,p_2\right).
\end{align}
This approximation is significant as it simplifies the expression by focusing on the most relevant components for our analysis. Consequently, the cross section can be further simplified to
\begin{align}
\sigma \simeq &\frac{(Ze)^{4}}{(4vk^0_1k^0_2)^2}\int\frac{\mathrm{d}^{2}\boldsymbol{b}_{\perp}\mathrm{d}^{3}\boldsymbol{p}_{1}\mathrm{d}^{3}\boldsymbol{p}_{2}}{(2\pi)^{2}E_{1}E_{2}} \notag \\
&\times \biggl|\int\frac{\mathrm{d}^{2}\boldsymbol{k}_{1\perp}\mathrm{d}^{2}\boldsymbol{k}_{2\perp}}{(2\pi)^{2}(2\pi)^{2}}\delta^{(2)}(\boldsymbol{k}_{1\perp}+\boldsymbol{k}_{2\perp}-\boldsymbol{p}_{1\perp}-\boldsymbol{p}_{2\perp})\mathcal{M}_{ij}(k_{1},k_{2},p_{1},p_{2})e^{i\bm k_{1\perp}\cdot \bm b_{\perp}}\boldsymbol{k}_{1\perp}^{i}\boldsymbol{k}_{2\perp}^{j}\frac{F(-k_{1}^{2})}{-k_{1}^{2}}\frac{F(-k_{2}^{2})}{-k_{2}^{2}}\biggr|^{2}.
\end{align}

Finally, we present the joint $b_\perp$ and $q_\perp$ dependent cross section for the process $A_{1}A_{2}\rightarrow A_{1}A_{2}\tau^{+}\tau^{-}$, as shown in Eq.~\eqref{eq:born}:
\begin{align}
\frac{\mathrm{d} \sigma}{\mathrm{d}^{2} \bm q_{\perp} \mathrm{d}^{2} \bm P_{\perp} \mathrm{d} y_{1} \mathrm{d} y_{2} \mathrm{d}^{2} \bm b_{\perp}} &= \notag \\
& \hspace{-1cm} \frac{\alpha_e^2}{2M^4\pi^2} \left[A_0+B_0^{(1)}F_2+B_0^{(2)} F_2^2 +C_0^{(2)} F_3^2+\left(A_2+B_2^{(2)}F_2^2+C_2^{(2)}F_3^2\right) \cos 2 \phi+A_4\cos 4\phi \right],
\end{align}
where $\bm P_{\perp} \equiv (\bm p_{1 \perp}-\bm p_{2 \perp})/2$, $y_{1,2}$ are rapidities of the leptons, $M$ is the invariant mass of the lepton pair, $\alpha_e$ is the fine-structure constant, and $\phi$ is the angle between transverse momentum $\bm q_{\perp}$ and $\bm P_{\perp}$. This expression encapsulates contributions from all interactions detailed in Eq.~\eqref{eq:eff_int}. Specifically, the terms $A_0, A_2,$ and $A_4$ represent the standard QED contributions to the process. On the other hand, the $B$ and $C$ terms are associated with the dipole interactions $F_2$ and $F_3$, respectively. Because the orthogonality of the polarization vectors of incoming photons with different helicity states allows $F_2$ and $F_3$ to induce specific angular modulations in the cross section. This phenomenon is reflected in the structure of the equation. For a complete understanding of the contribution of each term to the cross section, we present the full expressions of $A_0, A_2, A_4, B_{0}^{(1),(2)}, C_{0}^{(2)}, B_{2}^{(2)},$ and $C_{2}^{(2)}$ as follows.

\begin{eqnarray*}
A_{0} & = &\frac{1}{\left(m_{\tau}^2+P_{\perp}^2\right)^2}{\cal \int}[{d\cal K}_\perp] \Bigl[-2 m_{\tau}^4 \cos (\phi_{k_{1\perp}}+\phi_{\bar k_{1\perp}}-\phi_{k_{2\perp}}-\phi_{\bar k_{2\perp}})\\
&  &+P_{\perp}^2\left( M^2-2 P_{\perp}^2\right) \cos (\phi_{k_{1\perp}}-\phi_{\bar k_{1\perp}}+\phi_{k_{2\perp}}-\phi_{\bar k_{2\perp}}) +m_{\tau}^2 \left(M^2-2 m_{\tau}^2\right) \cos (\phi_{k_{1\perp}}-\phi_{\bar k_{1\perp}}-\phi_{k_{2\perp}}+\phi_{\bar k_{2\perp}})\Bigr],\\
A_2 & = &\frac{8m_{\tau}^{2}P_{\perp}^2}{\left(m_{\tau}^2+P_{\perp}^2\right)^2}{\cal \int}[{d\cal K}_\perp] \cos\left(\phi_{k_{1\perp}}-\phi_{k_{2\perp}}\right)\cos(\phi_{\bar k_{1\perp}}+\phi_{\bar k_{2\perp}}-2\phi_{q_\perp}),\\
A_4 & = &-\frac{2 P_{\perp}^4}{\left(m_{\tau}^2+P_{\perp}^2\right)^2}{\cal \int}[{d\cal K}_\perp] \cos (\phi_{k_{1\perp}}+\phi_{\bar k_{1\perp}}+\phi_{k_{2\perp}}+\phi_{\bar k_{2\perp}}-4 \phi_{q_\perp}),\\
B_{0}^{(1)} & = & \frac{2 M^2}{(m_{\tau}^2+P_{\perp}^2)^2} {\cal \int}[{\mathrm{d}\cal K}_\perp] \Bigl[-\left(m_{\tau}^2+P_{\perp}^2\right) \cos (\phi_{k_{1\perp}}+\phi_{\bar k_{1\perp}}-\phi_{ k_{2\perp}}-\phi_{\bar k_{2\perp}})\\
&  &+ P_{\perp}^2 \cos (\phi_{k_{1\perp}}-\phi_{\bar k_{1\perp}}+\phi_{ k_{2\perp}}-\phi_{\bar k_{2\perp}})+m_{\tau}^2 \cos (\phi_{k_{1\perp}}-\phi_{\bar k_{1\perp}}-\phi_{ k_{2\perp}}+\phi_{\bar k_{2\perp}})\Bigr], \\
B_{0}^{(2)} & = & \frac{M^2}{2 m_{\tau}^2 (m_{\tau}^2 + P_{\perp}^2)^2}
{\cal \int}[{\mathrm{d}\cal K}_\perp]
\Bigl[-5 m_{\tau}^2 \left(m_{\tau}^2+P_{\perp}^2\right) \cos (\phi_{k_{1\perp}}+\phi_{\bar k_{1\perp}}-\phi_{ k_{2\perp}}-\phi_{\bar k_{2\perp}})\\
&  & +2 P_{\perp}^2 \left(2 m_{\tau}^2+P_{\perp}^2\right) \cos (\phi_{k_{1\perp}}-\phi_{\bar k_{1\perp}}+\phi_{ k_{2\perp}}-\phi_{\bar k_{2\perp}}) +\left(5 m_{\tau}^4+5 m_{\tau}^2 P_{\perp}^2+2 P_{\perp}^4\right) \cos (\phi_{k_{1\perp}}-\phi_{\bar k_{1\perp}}-\phi_{ k_{2\perp}}+\phi_{\bar k_{2\perp}}) \Bigr],\\
C_{0}^{(2)} & = & \frac{M^2}{2 m_{\tau}^2 (m_{\tau}^2 + P_{\perp}^2)}
{\cal \int}[{\mathrm{d}\cal K}_\perp]
\Bigl[5 m_{\tau}^2 \cos (\phi_{k_{1\perp}}+\phi_{\bar k_{1\perp}}-\phi_{ k_{2\perp}}-\phi_{\bar k_{2\perp}})\\
&  &+2 P_{\perp}^2 \cos (\phi_{k_{1\perp}}-\phi_{\bar k_{1\perp}}+\phi_{ k_{2\perp}}-\phi_{\bar k_{2\perp}})+\left(3 m_{\tau}^2+2 P_{\perp}^2\right) \cos (\phi_{k_{1\perp}}-\phi_{\bar k_{1\perp}}-\phi_{ k_{2\perp}}+\phi_{\bar k_{2\perp}})\Bigr],\\
B_{2}^{(2)} & = & C_{2}^{(2)} = -\frac{2M^2 P_{\perp}^2}{m_{\tau}^2 (m_{\tau}^2 + P_{\perp}^2)}
{\cal \int}[{\mathrm{d}\cal K}_\perp]
\cos\left(\phi_{k_{1\perp}}-\phi_{k_{2\perp}}\right)\cos(\phi_{\bar k_{1\perp}}+\phi_{\bar k_{2\perp}}-2\phi_{q_\perp}).
\end{eqnarray*}
Here the shorthanded notation for the integration is defined as follows,
\begin{align}
{\cal \int}[\mathrm{d}{\cal K}_\perp] \equiv &\,\int \mathrm{d}^{2}\bm k_{1\perp}\mathrm{d}^{2}\bm k_{2\perp}\mathrm{d}^{2}\bar {\bm k}_{1\perp}\mathrm{d}^{2}\bar {\bm k}_{2\perp}e^{i(\bm k_{1\perp}-\bar {\bm k}_{1\perp})\cdot \bm b_{\perp}} \notag \\
&\times \delta^{(2)}(\bm k_{1\perp}+\bm k_{2\perp}-\bm q_{\perp}) \delta^{(2)}(\bar {\bm k}_{1\perp}+\bar {\bm k}_{2\perp}-{\bm q}_{\perp})  \mathcal{F}(x_1,{\bm k}_{1\perp}^{2})\mathcal{F}(x_2, {\bm k}_{2\perp}^{2})\mathcal{F}(x_1, \bar {\bm k}_{1\perp}^{2})\mathcal{F}(x_2, \bar {\bm k}_{2\perp}^{2}),
\end{align}
where the transverse momenta of the initial photons in the amplitude and its conjugate amplitude are denoted by $\bm k_{1,2\perp}$ and $\bar {\bm k}_{1,2\perp}$, respectively. The photon’s probability amplitude function ${\cal F}(x,{\bm k}_\perp^2)$, defined in Eq. \eqref{eq:gamma_amp}, is determined by the nuclear charge density distribution $F(-k^2)$. Besides, $x_1$ and $x_2$ represent the longitudinal momentum fractions of the incoming photons, given in Eq. \eqref{eq:energy_frac}.

\end{document}